\documentclass[aps, 
	prl, 
	noeprint, 
	superscriptaddress, 
	floatfix, 
	tightenlines, 
	twocolumn
	]{revtex4-1}
\usepackage[english]{babel}
\usepackage{graphicx}                      
\usepackage{listings} 
\usepackage{amsmath,amssymb,amsthm}        
\usepackage[utf8]{inputenc}
\usepackage[toc,page]{appendix}
\usepackage{color}
\usepackage{caption}

\usepackage{hyperref}
\hypersetup{
  colorlinks   = true, 
  urlcolor     = blue, 
  linkcolor    = blue, 
  citecolor    = blue 
}

\newcommand{\ket}[1]{\left | #1 \right \rangle}	

\begin{document}

\title{Entanglement between a telecom photon and an on-demand multimode solid-state quantum memory}
\author{Jelena V. Rakonjac}
\thanks{These two authors contributed equally.}
\affiliation{ICFO-Institut de Ciencies Fotoniques, The Barcelona Institute of Science and Technology, 08860 Castelldefels (Barcelona), Spain.}
\author{Dario Lago-Rivera}
\thanks{These two authors contributed equally.}
\affiliation{ICFO-Institut de Ciencies Fotoniques, The Barcelona Institute of Science and Technology, 08860 Castelldefels (Barcelona), Spain.}
\author{Alessandro Seri}
\affiliation{ICFO-Institut de Ciencies Fotoniques, The Barcelona Institute of Science and Technology, 08860 Castelldefels (Barcelona), Spain.}
\author{Margherita Mazzera}
\affiliation{Institute of Photonics and Quantum Sciences, SUPA, Heriot-Watt University, Edinburgh EH14 4AS, UK.}
\author{Samuele Grandi}\email[]{samuele.grandi@icfo.eu}
\affiliation{ICFO-Institut de Ciencies Fotoniques, The Barcelona Institute of Science and Technology, 08860 Castelldefels (Barcelona), Spain.}
\author{Hugues de Riedmatten}
\affiliation{ICFO-Institut de Ciencies Fotoniques, The Barcelona Institute of Science and Technology, 08860 Castelldefels (Barcelona), Spain.}
\affiliation{ICREA-Instituci\'o Catalana de Recerca i Estudis Avan\c cats, 08015 Barcelona, Spain.}

\begin{abstract}
Entanglement between photons at telecommunication wavelengths and long-lived quantum memories is one of the fundamental requirements of long-distance quantum communication. Quantum memories featuring on-demand read-out and multimode operation are additional precious assets that will benefit the communication rate. In this work we report the first demonstration of entanglement between a telecom photon and a collective spin excitation in a multimode solid-state quantum memory. Photon pairs are generated through widely non-degenerate parametric down-conversion, featuring energy-time entanglement between the telecom-wavelength idler and a visible signal photon. The latter is stored in a Pr$^{3+}$:Y$_2$SiO$_5$ crystal as a spin wave using the full Atomic Frequency Comb scheme. We then recall the stored signal photon and analyze the entanglement using the Franson scheme. We measure conditional fidelities of 92(2)\% for excited-state storage, enough to violate a CHSH inequality, and 77(2)\% for spin-wave storage. Taking advantage of the on-demand read-out from the spin state, we extend the entanglement storage in the quantum memory for up to 47.7~$\mu$s, which could allow for the distribution of entanglement between quantum nodes separated by distances of up to 10~km.

\end{abstract}

\maketitle


At the basis of the operation of a quantum network lies the ability to distribute entanglement between remote locations \cite{Duan2001}, and several examples have already been demonstrated \cite{Chou2005,Chou2007,Moehring2007,Yuan2008,Ritter2012,Hofmann2012,Usmani2012,Delteil2016,Sipahigil2016,Humphreys2018,Puigibert2020,Yu2020,Lago-Rivera2021,Liu2021}. Any physical implementation would have to encompass high storage efficiencies and entanglement generation rates. Addressing communication over long distances requires the preservation of light-matter entanglement over long storage times, ideally with on-demand retrieval. At the same time, a direct integration with the telecom fiber network would provide a natural landscape for the deployment of a quantum network, triggering the need for entanglement between quantum memories and photons at telecommunication wavelengths where the loss in fiber is minimal. Finally, multiplexing is a fundamental resource for achieving practical rates over long-distances \cite{Sangouard2011}.

Atomic systems offer a natural way of generating light-matter entanglement: in atomic clouds \cite{Matsukevich2005,deRiedmatten2006,Inoue2006,Chen2007a,Dabrowski2017,Farrera2018} as well as in single neutral atoms \cite{Volz2006,Wilk2007}. Telecom compatibility can be added using suitable excitation schemes \cite{Chang2019} or frequency conversion \cite{Radnaev2010,VanLeent2020}. Similarly, single trapped ions offer a promising platform \cite{Walker2018,Bock2018,Krutyanskiy2019}, with long storage times and the possibility of performing local operations, which led to the demonstration of light-matter entanglement over up to 50~km of optical fiber. Solid-state alternatives, such as quantum dots \cite{Gao2012,DeGreve2012} or color centers in diamond \cite{Tchebotareva2019}, have also been explored due to their potential for scalability \cite{Kalb2017}. Multimode operation is however limited to only a few systems \cite{Inoue2006,Chang2019}, with rare-earth systems providing a promising solution \cite{Clausen2011,Saglamyurek2011}.

We present here our proposal for light-matter entanglement distribution, built on a long-lived, solid-state and multimode quantum memory paired with an external source of entangled photon pairs \cite{Simon2007}. This source is based on a parametric down-conversion crystal placed in an optical cavity \cite{Fekete2013}. The down-conversion process is highly non-degenerate: while the idler photon is in the telecom E-band, the signal is generated at $606$~nm. The optical cavity shapes the spectral profile of the generated photons, ensuring compatibility with a solid-state quantum memory \cite{Rielander2014, Seri2018}, in our case a Y$_2$SiO$_5$ crystal doped with praseodymium (Pr) ions. The signal photons are stored as a collective excitation of the ions through the full Atomic Frequency Comb protocol \cite{Afzelius2009} that allows on-demand readout of the photon. Pr-based memories have already shown great potential: the storage of time and polarization qubits \cite{Gundogan2012, Gundogan2015} and the generation of light-matter entanglement \cite{Ferguson2016,Kutluer2017, Kutluer2019} have been demonstrated in this system, as well as matter-matter entanglement \cite{Lago-Rivera2021,Liu2021}. By pairing this quantum memory with our external photon source we obtain a system that allows for time \cite{Seri2017}, frequency \cite{Rielander2018, Seri2019} and spatial \cite{Seri2018, Yang2018} multiplexing. In this work we push its reach forward by demonstrating energy-time entanglement between the telecom idler photon and the visible signal one stored as a delocalized excitation of the ions' spin ground state for tens of microseconds.

\begin{figure*}
	\centering
	\includegraphics[width=2\columnwidth]{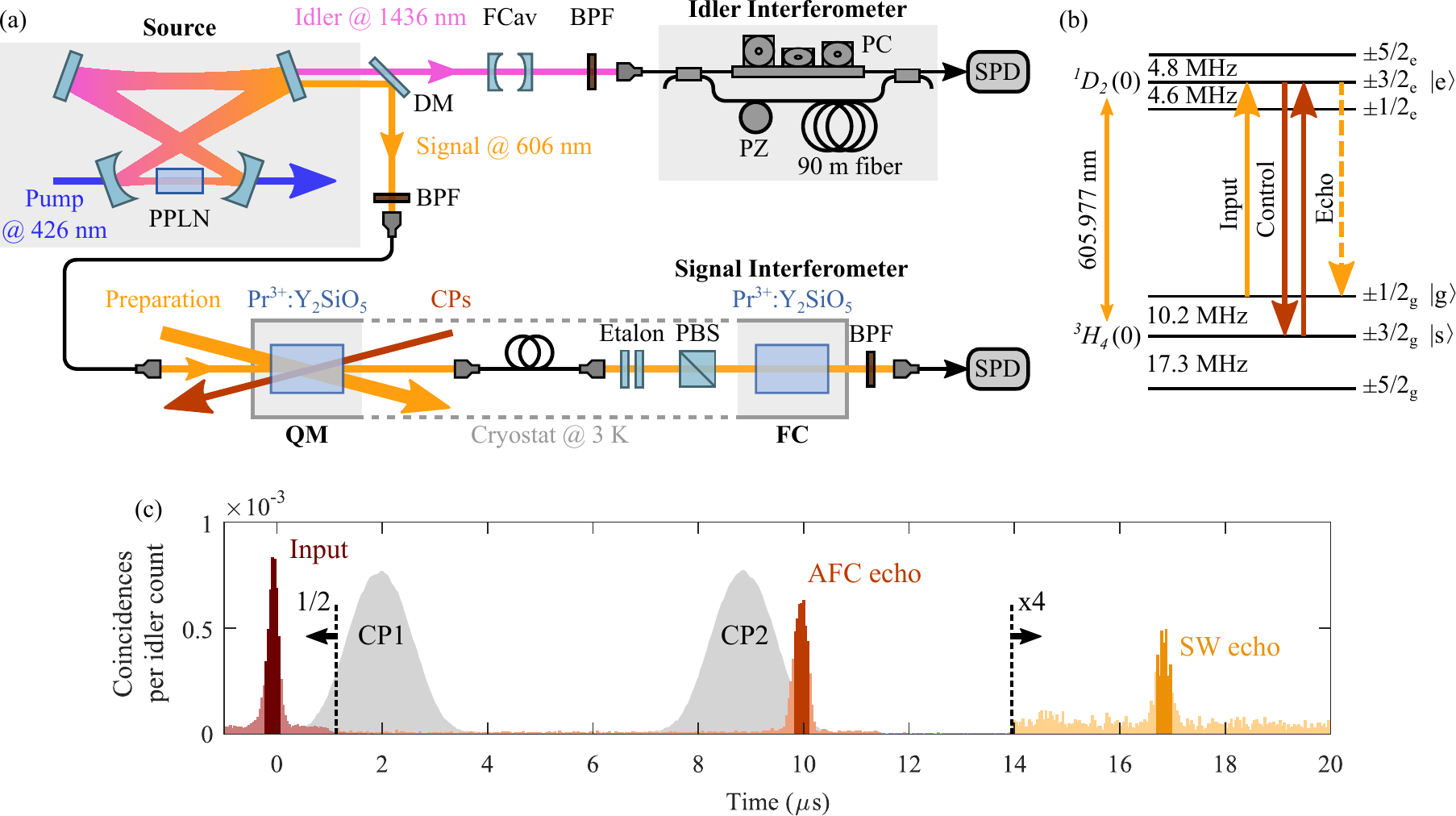}
	\caption{(a) Experimental set-up. Entangled photon pairs generated at the source are separated with a dichroic mirror (DM). Telecom photons are sent directly to an interferometer for entanglement analysis. Signal photons are stored in the quantum memory crystal (QM), and are later filtered and analyzed in the filter crystal (FC). PPLN: periodically-poled lithium niobate; FCav: Fabry-Perot filter cavity; BPF: band-pass filter; PZ: piezo actuator; PC: polarization controller; SPD: single-photon detector. (b) Energy level scheme of Pr$^{3+}$:Y$_2$SiO$_5$, where the relevant transitions are indicated. (c) Coincidence histograms showing the time difference between the detection of idler and signal photons. The darker regions indicate the portion considered for analysis. We plot the three cases that we analyze in the text: input photons, AFC storage for 10 $\mu$s and SW storage with a $T_s$ of 6.9 $\mu$s.}
	\label{fig:setup}
\end{figure*}

Energy-time entangled photon pairs are generated using a cavity-enhanced spontaneous parametric down conversion (cSPDC) source (Fig.~\ref{fig:setup}(a)). We pump a periodically-poled lithium niobate crystal with $4$ mW of continuous-wave light at 426 nm. Only photon pairs in resonance with the optical cavity modes are created: the idler, at a telecom wavelength (1436 nm), and the signal (606 nm), in resonance with the $^3$H$_4$(0)$\leftrightarrow$ $^1$D$_2$(0) transition of Pr$^{3+}$:Y$_2$SiO$_5$.
The generation of energy-time entanglement is ensured as long as $\tau_\textit{pump} \gg \tau_\textit{pair}$, where $\tau_\textit{pump}$ ($\tau_\textit{pair}$) corresponds to the coherence time of the pump laser light (photon pair). In our case, $\tau_\textit{pair}= 120$~ns while, by assuming a gaussian distribution of phase noise, we estimate that $\tau_\textit{pump}\backsimeq1$~$\mu$s \cite{SuppMat}.
The spectrum of the photon pair is composed of 15 frequency modes separated by 261.1~MHz \cite{Seri2019}; in order to select a single frequency mode the idler photons are spectrally filtered with a Fabry-Perot filter cavity. As explained below, a single frequency mode of the signal photons is selected directly by the quantum memory. In addition, we switch off the pump laser 1 $\mu$s after detecting an idler photon and we keep it off for few tens of microseconds to prevent the creation of additional pairs during the retrieval of the stored photons.

\begin{figure}[hbtp]
	\centering
	\includegraphics[width=\columnwidth]{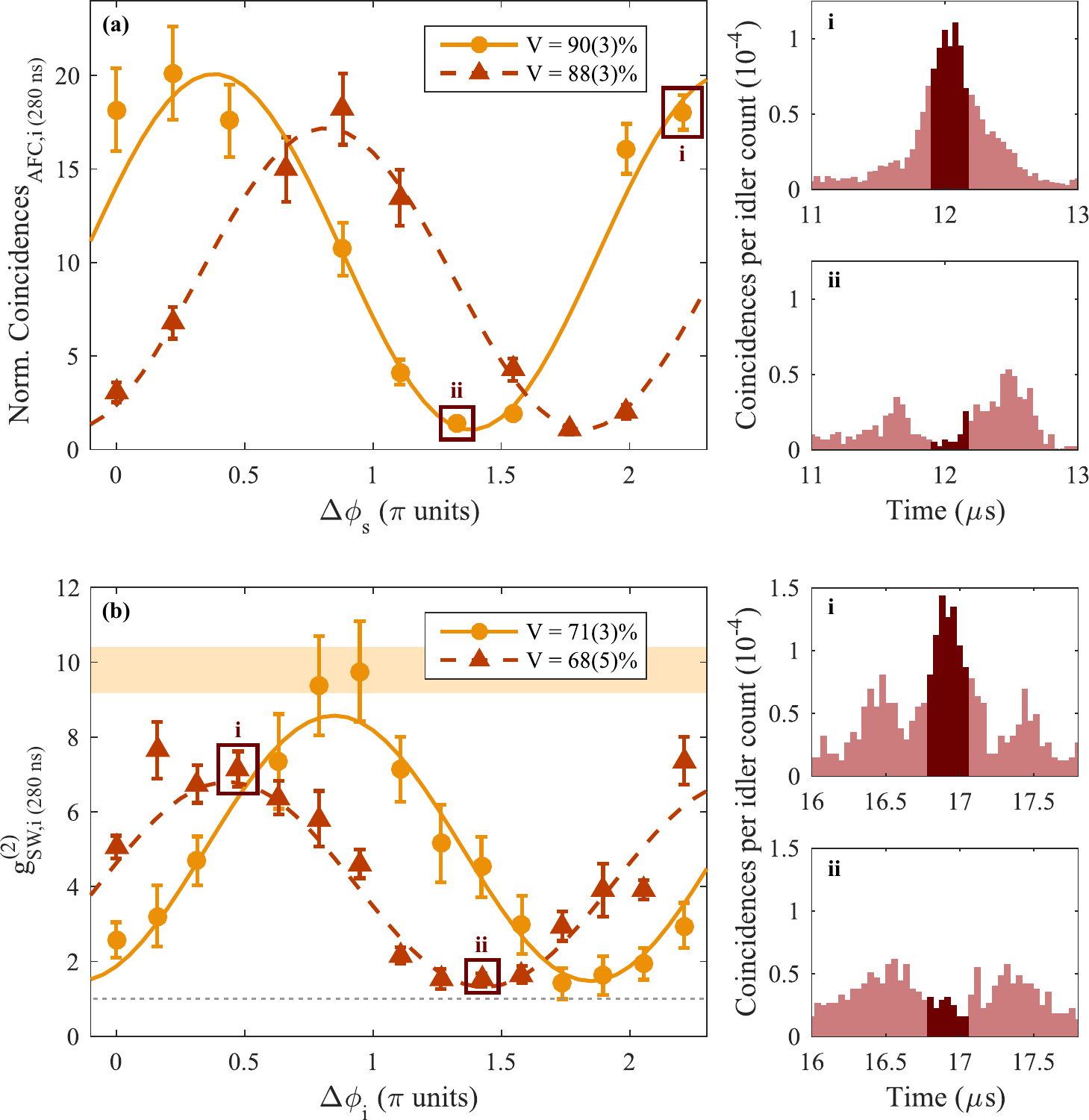}
	\caption{Visibility fringes and relevant coincidence histograms for maximum and minimum values (darker regions represent the coincidences window used). (a) AFC storage. We varied the phase between the short and long paths of the AFC-based interferometers for two settings of the idler interferometer differing by $\pi$/2. See \cite{SuppMat} for the normalization of the AFC coincidences. (b) Semiconditional SW storage. We scanned the phase between the short and long paths of the idler interferometer for two settings of the AFC-based interferometer differing by $\pi$/2. The shaded area represents the maximum value that we expect for the ideal case, with a span of 1 sigma, while the dashed-grey line is the minimum one.}
	\label{fig:fringes}
\end{figure}

We perform spin-wave storage of the signal photons using the full AFC scheme in a Pr$^{3+}$:Y$_2$SiO$_5$ quantum memory (QM) crystal \cite{Afzelius2009}. We prepare an AFC by tailoring the absorption spectrum of the 1/2$_g$~-~3/2$_e$ transition (Fig.~\ref{fig:setup}(b)) into a periodical structure while keeping the 3/2$_g$ spin level empty.
An incoming photon is then absorbed by the ions and is mapped to a collective atomic excitation. After an initial dephasing the atoms will rephase at a predefined storage time $\tau_\textit{AFC}$, given by the inverse of the periodicity of the comb, therefore mapping the excitation back into a light field, called the AFC echo. Before this re-emission occurs, we send a control pulse (CP - see \cite{SuppMat} for more details) that coherently drives the collective excitation to the ground-state level 3/2$_g$. This stops the collective rephasing for an arbitrary time $T_s$, until a second CP retrieves the excitation back to the 3/2$_e$ level. The AFC echo resulting from the storage in the spin state is called the spin-wave echo. Since $T_s$ can be varied within the coherence time of the spin levels, spin-wave storage allows for on-demand readout of the QM.
However, the CPs lead to the generation of uncorrelated light, through laser light coupling into the detection spatial mode, fluorescence and free induction decay originating from residual population in the $3/2_g$ state. To minimize this noise, we performed spatial filtering by introducing a small angle between the optical path of the signal photons and of the CPs, which are also counter-propagating with respect to the signal.
We also used a second Pr$^{3+}$:Y$_2$SiO$_5$ crystal, referred to as Filter Crystal (FC), as an ultra-narrowband spectral filter. We prepared a 6 MHz wide transparency window centered at the frequency of the AFC using spectral hole-burning, with an optical depth of 6 outside of it. Additionally, we placed an etalon filter in between the crystals and a band-pass filter before the detector to remove the broadband noise that is not filtered by the inhomogeneous broadening of the QM and the FC. Further filtering is provided by a polarization beam splitter between the QM and the FC that removes part of the unpolarized fluorescence (Fig. \ref{fig:setup}(a)).

First, we characterized the quality of the source and of the quantum memory. We measured correlations between the idler photons and the signal photons, and used the second-order cross-correlation function as a figure of merit. This is defined as $g^{(2)}_{s,i(\Delta t)} = \frac{p_\textit{s,i}}{p_s\cdot p_i}$, where $p_\textit{s,i}$ represents the probability of detecting a coincidence between an idler and a signal photon in a time window $\Delta t$, and $p_s$ ($p_i$) is the unconditional probability of detecting a signal (idler) photon in the same window. Histograms with raw coincidences are reported in Fig.~\ref{fig:setup}(c). We first prepared a 20~MHz wide transparency window in the QM, and measured $g^{(2)}_\textit{s,i (280ns)} = 23.3(2)$, well above the classical limit of 2 (assuming thermal statistics for the two fields).
Next, we stored the signal photons in the QM with an AFC with a storage time of 10~$\mu$s. We measured $g^{(2)}_\textit{AFC,i (280ns)} = 91(4)$ with an efficiency of $\eta_\textit{AFC} = 19.7(5) \%$. The increase in $g^{(2)}$ with respect to the previous measurement is a consequence of the additional spectral and temporal filtering that the AFC storage provides \cite{Rielander2014,Seri2017}.
Finally, we performed spin-wave storage by sending two CPs separated by 6.9~$\mu$s, conditioned on the detection of an idler photon. As they are the main source of noise in our measurement, we subsequently sent N pairs of CPs every 400~$\mu$s, and averaged the noise level at the position where a spin echo would have been over all these trials. The value of \textit{N} is optimized for each case so as to minimize the measuring time required to obtain a significant error on the $g^{(2)}$, while the 400~$\mu$s separation was chosen as to minimize the residual fluorescence from the CPs after every storage trial. We call this method semiconditional spin-wave storage \cite{Seri2017}. We measured a noise floor of $8.3(4) \cdot 10^{-4}$ photons per storage trial, resulting in a cross-correlation of $g^{(2)}_\textit{SW,i (280ns)} = 9.8(6)$ with an efficiency of $\eta_\textit{SW} = 6.2(3) \%$. Higher values can be attained, at the cost of an increase in noise \cite{SuppMat}. Considering the AFC storage time and the control pulse width of 3~$\mu$s, our system allows the storage of 17 independent temporal modes with a width of 420~ns.
The increase of $g^{(2)}$ compared to our previously published results \cite{Seri2017} was due to improvements in the AFC preparation, noise filtering, and better spectral overlap with the source \cite{SuppMat}.

We then moved on to measure the quality of the entanglement storage in our quantum memory. To analyze the entanglement, we followed the scheme suggested by Franson \cite{Franson1989} that involves sending both signal and idler photons through unbalanced Mach-Zehnder interferometers. This post-selects the state $(\ket{EE} + e^{i \phi}\ket{LL})/\sqrt{2}$ for signal and idler, where $\ket{E}$ and $\ket{L}$ are time bins associated to the short and long arm of the interferometer respectively, and $\phi$ depends on the relative phase between the two arms. Therefore, the time difference $\tau_{MZ}$ between them has to fulfill the condition $\tau_\textit{pump}> \tau_\textit{MZ} > \tau_\textit{pair}$. In our case we chose $\tau_\textit{MZ} = 420$~ns. More importantly, this scheme requires a quantum memory supporting multimodality, since it involves the storage of (at least) two temporal modes. We used a fiber-based interferometer for the idler photon and an AFC-based one for the signal photon \cite{Clausen2011,Maring2017,Kutluer2019}.

We started by analyzing storage of entanglement in the excited state of the quantum memory. We prepared two different AFCs in the filter crystal, in such a way that a signal photon could be stored for 2~$\mu$s (short path) or 2.42~$\mu$s (long path) with equal probability. Moving the central frequency of one AFC with respect to the other adds a phase shift between the short and long paths. We scanned this phase while keeping that of the idler interferometer constant in order to reconstruct an interference fringe. We then shifted the phase of the idler interferometer by $\pi/2$ to change the measurement basis and we repeated the same measurement. The results are reported in Fig.~\ref{fig:fringes}(a); we obtained two visibility values of 90(3)\% and 88(3)\%, respectively 6 and 5 standard deviations above the threshold of 70.7\% required to violate a Clauser-Horne-Shimony-Holt (CHSH) inequality. These visibilities {\itshape $V$} lead to an average two-qubit conditional fidelity of 92(2)\% for the light-matter entanglement \cite{Clauser1969} (calculated as $(3V + 1)/4$, assuming white noise).

We then moved to spin-wave storage of the signal photons. For this case, we prepared in the filter crystal a single AFC with a storage time of 420~ns. The long and the short path of the interferometer are represented, respectively, by storage in the AFC and by the transmission through it, without absorption \cite{Jobez2015a} (see \cite{SuppMat} for a discussion on the advantages of the different methods). We then scanned the phase of the idler interferometer for two different phase settings of the solid-state interferometer and recorded the interference fringes. The results are reported in Fig.~\ref{fig:fringes}(b) and correspond to visibilities of 71(3)\% and 68(5)\%, leading to a two-qubit conditional fidelity of 77(2)\%. Even though these values do not exceed the CHSH limit, the visibilities are above the 33\% bound for separable states by 12 and 7 standard deviations, respectively \cite{Peres1996}. This is the first demonstration of entanglement between a telecom photon and a solid-state on-demand multimode quantum memory.

\begin{figure}[tp]
	\centering
	\includegraphics[width=\columnwidth]{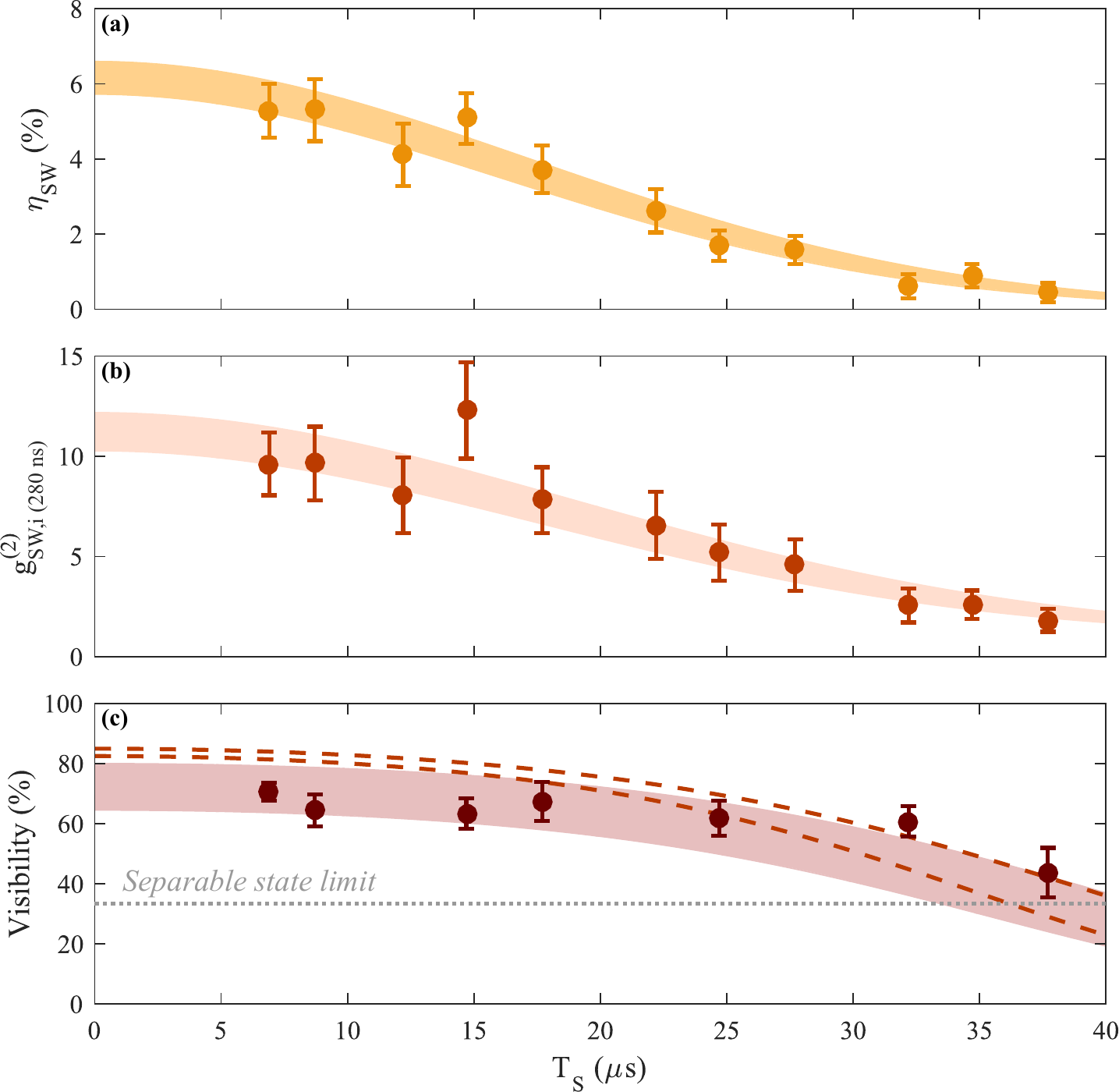}
	\caption{(a, b) Variation of the storage efficiency and second-order cross-correlation with $T_s$. They follow a Gaussian trend, which we use for the fit, obtaining an inhomogeneous broadening of the spin state of 16.1(7) and 14.8(9) kHz, respectively. The shaded area corresponds to one standard deviation of the error of the fit. (c) Variation of the entanglement visibility with $T_s$. The shaded area represents our model accounting for imperfect analyzers, while the dashed lines correspond to the upper and lower bounds of the ideal case, only limited by photon statistics. The dotted gray line represents the bound for separable states.}
	\label{fig:Tsdep}
\end{figure}

The key advantage of this storage is the ability to retrieve the stored photon in an on-demand fashion; by varying the time between the CPs we can then investigate the effect of $T_s$ on the quality of the correlations and of the entanglement.
We first measured the efficiency $\eta_\textit{SW}$ and $g^{(2)}_\textit{SW,i}$ as a function of $T_s$ (Fig. \ref{fig:Tsdep}(a, b)), and observed that they decreased with increasing $T_s$. This is to be expected, due to the inhomogeneous broadening of the spin state. We fitted the efficiency data to the expression $\eta_\textit{SW} = a_{\eta}\cdot e^{-\frac{(T_s\cdot \gamma_\textit{inhom}\cdot \pi)^2}{2\cdot \log(2)}}$, where $a_\eta$ depends on $\eta_\textit{AFC}$ and on the efficiency of the control pulses, and obtained a spin state inhomogeneous broadening $\gamma_\textit{inhom}$ of 16.1(7)~kHz \cite{Afzelius2010a,Gundogan2015,Seri2017}, which corresponds to a $1/e$ decay time of 23(1)~$\mu$s. We fitted the $g^{(2)}_\textit{SW,i}$ data to a similar function, and obtained a decay of 25(1)~$\mu$s. For more details regarding the derivation of the dependence of the $g^{(2)}_\textit{SW,i}$ as a function of $T_s$, see \cite{SuppMat}.
In addition, we measured the change in entanglement visibility as a function of $T_s$. The results are reported in Fig.~\ref{fig:Tsdep}(c).
To model these data, we first plotted the variation of visibility with $T_s$ in the case where it is only limited by photon statistics, i.e., we calculated the expected values of visibility according to $V = \frac{g^{(2)}_\textit{SW,i}-1}{g^{(2)}_\textit{SW,i}+1}$ \cite{deRiedmatten2006}, using values of $g^{(2)}_\textit{SW,i}$ extracted from the fit to the data of Fig.~\ref{fig:Tsdep}(b). The result is plotted in Fig.~\ref{fig:Tsdep}(c), with the dashed lines representing the upper and lower bounds of the visibility that we expect from this scenario.
We then developed a new model which accounts for the effect of spin inhomogeneity on imperfect interference \cite{SuppMat}. This is a more complete description of the physics at play, as it accounts for imperfect analyzers. We fit this new model to our data and we obtained the shaded area in Fig.~\ref{fig:Tsdep}(c), that provides a better fit to the experimental data. We could then conclude that for all the investigated values of $T_s$ the visibility was higher than the classical bound of 33\% \cite{Peres1996}, demonstrating storage of entanglement up to 37.7~$\mu$s in the spin-state of our on-demand QM and for a total storage time of 47.7 $\mu$s.

We have here reported the first demonstration of entanglement between a photon in the telecommunication band and an on-demand multimode quantum memory. We explored various scenarios using the memory as a predetermined delay line with an AFC of 10 $\mu$s and as an on-demand quantum memory, varying the time in the spin state from 6.9~$\mu$s to 37.7~$\mu$s. We demonstrated entanglement between both parties for all cases, up to a total storage time of almost 50~$\mu$s, which would allow enough time for the photon to propagate 10 km in an optical fiber. Moreover, in the case of excited-state storage, our system allows for a violation of a CHSH inequality. Thanks to temporal multimodality of the AFC protocol and to the 10~$\mu$s AFC storage time, our system would allow an increase in entanglement rate of a factor of 17 with respect to a similar one with single-mode operation, or half of this for storage of time-bin entanglement. Spatial and frequency multiplexing, already demonstrated in our system \cite{Seri2018, Seri2019}, could provide a further boost. In its current state, the visibility of the entanglement fringes is limited by the $g^{(2)}_\textit{SW,i}$ of the recalled spin-wave, besides from the efficiency of the entanglement analyzers. The $g^{(2)}_\textit{SW,i}$ could be increased by improving the noise filtering, for example by using the filter crystal in double-pass configuration, and by increasing the storage efficiency of the quantum memory. A careful shaping of the spectral and temporal profile of the control pulses could improve both signal-to-noise and storage efficiency \cite{Jobez2016} and the former could benefit from cavity-enhanced memories \cite{Afzelius2010b}.
Finally, applying spin-echo and dynamical decoupling techniques \cite{Heinze2013,Laplane2017,Holzapfel2020,Ma2020} will open the door to applications in quantum repeaters. The storage time of Pr-memories could thus be extended to tens of milliseconds, allowing the telecom photons to travel through tens of kilometers in optical fibers while the memory stores the entanglement, an essential steppingstone for long-distance quantum communication.

\section{Acknowledgments}
This project received funding from the European Union Horizon 2020 research and innovation program within the Flagship on Quantum Technologies through grant 820445 (QIA) and under the Marie Sk\l odowska-Curie grant agreement No. 713729 (ICFOStepstone 2) and No. 758461 (proBIST), from the European Union Regional Development Fund within the framework of the ERDF Operational Program of Catalonia 2014-2020 (Quantum CAT), from the Gordon and Betty Moore foundation through Grant GBMF7446 to HdR, from the Government of Spain (PID2019-106850RB-I00; Severo Ochoa CEX2019-000910-S; BES-2017-082464), from Fundaci\'o Cellex, Fundaci\'o Mir-Puig, and from Generalitat de Catalunya (CERCA, AGAUR).


\end{document}